\documentclass[aps,prb,twocolumn,floatfix]{revtex4}
\usepackage{graphicx}
\begin{document}
\title{Low-energy effective representation of the Gutzwiller-projected BCS Hamiltonian
close to half filling}
\author{Evgueny Kochetov}
\affiliation{Bogoliubov Theoretical Laboratory, Joint Institute for
Nuclear Research, 141980 Dubna, Russia\\ International Center of
Condensed Matter Physics, Universidade de Brasilia, Caixa Postal
04667, 70910-900 Brasilia, DF, Brazil}
\author{Alvaro Ferraz}
\affiliation{International Center for Condensed Matter Physics,
Universidade de Brasilia, Caixa Postal 04667, 70910-900 Brasilia,
DF, Brazil}
\author{Rafael T. Pepino}
\affiliation{International Center for Condensed Matter Physics,
Universidade de Brasilia, Caixa Postal 04667, 70910-900 Brasilia,
DF, Brazil}

\begin{abstract}
We investigate analytically a connection between the t-J model and the strongly correlated
Bardeen-Cooper-Schrieffer (BCS) Hamiltonian, with the effect of strong electron correlations accounted by the Gutzwiller projection.
We show that in the immediate vicinity of half filling the projected $2D$ BCS Hamiltonian with strong pairing
develops an antiferromagnetically (AF) ordered ground state. This result explicitly demonstrates that antiferromagnetism in this model appears as a natural consequence of the strong Coulomb repulsion in  a low doped regime.
At moderate doping
the ground state of the Gutzwiller-projected BCS Hamiltonian  becomes qualitatively similar to Anderson's resonating valence bond state which is known to fit nicely the properties of the t-J model in this regime.
These two properties taken together indicate that
the projected BCS Hamiltonian captures the essential low-energy physics of the t-J model in the whole underdoped region.

\end{abstract}
\pacs x 74.20.Mn, 74.20.-z
\maketitle

\section{Introduction}

The investigation of strongly correlated electron systems has been a central issue in solid state physics for more than four decades. The discovery of high-T$_c$ superconductivity in copper-oxide based compounds (cuprates) revived the interest in simple models displaying such strong correlations. Two much investigated models are the Hubbard model and its "descendant", the t-J model \cite{dagotto94,lee06}. One of the main theoretical questions in that field is whether or not there is a superconducting phase in the t-J model \cite{dagotto94}. Besides, the interplay between antiferromagnetism and superconductivity in the cuprates as well as their sensitivity to doping is still not very well understood.

It is clear that in superconducting state induced by electron-electron interaction the formation of Cooper pairs must also reflect strong electron correlations. As a result the BCS effective Hamiltonian should be directly modified by the inclusion of a non-double occupancy (NDO) constraint to account for such an effect.

In a recent paper, Park discussed a close connection between the t-J model and the Gutzwiller-projected BCS Hamiltonian \cite{park}. It was shown both numerically and analytically that the ground states of the t-J model at half filling (i.e. of the $2D$ aniferromagnetic Heisenberg model) and of the strongly correlated BCS hamiltonian are equivalent to each other.
Moreover, at sufficiently small doping, there is numerical evidence of a strong overlap between those two ground state wavefunctions, which provides further support for the existence of superconductivity in the t-J model. Clearly it would be interesting to establish by analytical means such an equivalence at non-zero hole concentration. As is known, slightly away from half filling
the long-range AF order is still observed in the cuprate superconductors.
If the projected BCS Hamiltonian is indeed believed to contain close to half filling the low-energy physics of the t-J Hamiltonian, its ground state must also exhibit the AF order in the immediate vicinity of half filling.
This manifests itself as a quite nontrivial necessary condition for the low-energy physics described by the Gutzwiller-projected BCS Hamiltonian to be considered identical to that of the t-J Hamiltonian at sufficiently low doping.

The purpose of the present report is to investigate the Gutzwiller-projected BCS Hamiltonian, close to half filling, analytically. We do not address here the issue of the properties of the t-J model at moderate doping, concentrating our full attention instead to the region of the phase diagram very close to half filling.
We derive the low-energy long-wavelength effective action for the lightly doped $2D$ projected BCS Hamiltonian on a bipartite lattice. The action obtained is shown to be identical to that of the $2D$ quantum antiferromagnetic Heisenberg model  explicitly represented by the $3D$ nonlinear $\sigma$-model. In other words, close to half filling the ground state of the Gutzwiller-projected BCS Hamiltonian is antiferromagnetically ordered and non-superconducting.
Since the conventional BCS Hamiltonian does not exhibit any magnetic ordering and always displays superconductivity, those results explicitly demonstrate that antiferromagnetism appears as a natural consequence of the strong Coulomb repulsion at low doped regimes.

Formally, the Gutzwiller projection takes care of the strong electron correlation due to the large on-site Coulomb repulsion. Close to half filling, the infinitely strong short-range Coulomb repulsion between lattice electrons brings about the superexchange of the emergent local spin moments by means of the virtual exchange processes which involve virtual creation of the electron spin singlets on the nearest-neighbor (nn) empty sites.
As a result the Gutzwiller-projected strong-pairing BCS Hamiltonian can be described in terms of the emergent spin-spin exchange interaction and the AF ordering continues from half filling up to a small doping.
In other words the projected BCS Hamiltonian can describe both the superconducting state at moderate doping and, in contrast with the conventional BCS Hamiltonian, the ordered magnetic phase for sufficiently low doping.

This paper is organized as follows. In Sec.II we set up the necessary notation and emphasize the importance of the Gutzwiller projection close to half filling. In Sec.III the low-energy action for the Gutzwiller-projected BCS Hamiltonian is derived within the coherent-state path-integral approach. We provide an independent operator derivation in Sec.IV. We conclude in Sec.V by discussing some physical implications of the obtained low-energy representation of the projected BCS Hamiltonian. Necessary technical details are discussed in Appendices.

\section{Gutzwiller-projected BCS Hamiltonian}
We start with the Gutzwiller projected BCS Hamiltonian on a $2D$
bipartite lattice, $L=A\oplus B$:
\begin{eqnarray}
H^{G}_{BCS}&=&\hat{\mathcal{P}}_GH_{BCS}\hat{\mathcal{P}}_G
\nonumber\\
&=&\hat{\mathcal{P}}_G (H_t+H_{\Delta}) \hat{\mathcal{P}}_G,
\label{1.1}\end{eqnarray} where $$H_t= -t\sum_{ij\sigma
}\left(c_{i\sigma}^{\dagger}c_{j\sigma}+H.c.\right)$$ is a kinetic
term of strength $t$ responsible for the hopping of electrons from
one lattice site to its nearest neighbor and
\begin{equation}
H_{\Delta}=\sum_{ij}\Delta_{ij}\left(c_{i\uparrow}^{\dagger}c^{\dagger}_{j\downarrow}
-c_{i\downarrow}^{\dagger}c^{\dagger}_{j\uparrow} +H.c.\right)
\label{1.2}
\end{equation}
is the pairing term in real space. Here $c_{i\sigma}$ is the
electron annihilation operator at site $i$ with the spin
projection $\sigma=\uparrow\downarrow$.

At every lattice site the Gutzwiller projection operator
$$\hat{\mathcal{P}}_G=\prod_i(1-n_{i\sigma}n_{i-\sigma}), \,n_{i\sigma}=
c^{\dagger}_{i\sigma}c_{i\sigma}$$ projects out the doubly
occupied states $|\uparrow\downarrow\rangle$ thereby reducing the
quantum Hibert space to a lattice site product of the
$3$-dimensional spaces spanned by $|0\rangle_i, \,
|\uparrow\rangle_i, \, |\downarrow\rangle_i.$ Physically this
modification of the original Hilbert space results in strong
electron correlation effects which are believed to account for
the unusual and rich physics of the high -T$_c$ superconductors.

Upon introducing a full set of the on-site operators
$X^{ab}:=|a\rangle\langle b|, \, a,b=0,\uparrow, \downarrow,$
which are also referred to as the Hubbard operators, the
Gutzwiller projection is explicitly evaluated to be
$$\hat{\mathcal{P}}_G c^{\dagger}_{i\sigma}\hat{\mathcal{P}}_G=
c^{\dagger}_{i\sigma}(1-n_{i-\sigma})=X_i^{\sigma 0},$$
$$\hat{\mathcal{P}}_Gn_i\hat{\mathcal{P}}_G\equiv \tilde n_i=n_i-2n_{i\uparrow}n_{i\downarrow}=
X_i^{\uparrow\uparrow}+X_i^{\downarrow\downarrow},$$ where
$X^{\uparrow\uparrow}+X^{\downarrow\downarrow}+X^{00}=|0\rangle\langle
0|+|\uparrow\rangle \langle\uparrow|+ |\downarrow\rangle\langle
\downarrow|=\hat I$ is the identity operator in the on-site
Gutzwiller-projected Hilbert space. Note that the eigenvalues of
the projected electron number operator, $\tilde n_i$, are either 0
or 1, so that the doubly occupied states are prohibited.
It should be stressed that it is close to half filling that the Gutzwiller projection is of a crucial importance: the projected electron operator $\hat{\mathcal{P}}_G c^{\dagger}_{i\sigma}\hat{\mathcal{P}}_G$ in this region significantly differs from the bare electron operator  $c^{\dagger}_{i\sigma}$ (right at half filling $\hat{\mathcal{P}}_G c^{\dagger}_{i\sigma}\hat{\mathcal{P}}_G=0$).

With these notations Eq.~(\ref{1.1}) can be rewritten in the
equivalent form,
\begin{eqnarray}
H^G_{BCS}&=&-t\sum_{ij\sigma }\left(X_{i}^{\sigma 0}X_{j}^{0\sigma
}+H.c.\right) +\mu\sum_iX_i^{00}\nonumber\\
&+&\sum_{ij}\Delta_{ij}\left(X^{\uparrow 0}_{i}X^{\downarrow
0}_{j} -X^{\downarrow 0}_{i}X^{\uparrow 0}_{j} +H.c.\right),
\label{1.3}
\end{eqnarray}
where we have introduced the chemical potential term to control
the total number of doped holes, $$X^{00}_i=1-\tilde
n_i=(1-n_i)^2.$$ The local NDO constraint is rigorously
taken into account at the expense of the introduction of the
Hubbard operators with more complicated commutation relations than
those of the standard fermion algebra. In fact, fermionic
operators $X_{i}^{\sigma 0}$ together with the bosonic ones,
$X^{\sigma\sigma'}_i$  form, on every lattice site, a basis of the
fundamental representation of the graded (supersymmetrical) Lie
algebra su$(2|1)$ given by the (anti)commutation relations
\begin{equation}
\{X^{ab}_i,X^{cd}_j\}_{\pm}=(X_i^{ad}\delta^{bc}\pm X^{bc}_j
\delta^{ad})\delta_{ij}, \label{1.4}\end{equation}
where the $(+)$ sign should be used only
when both operators are fermionic.

In the strong-pairing limit $(|\Delta|>> t)$ the projected BCS
hamiltonian (\ref{1.3}) reduces to
\begin{eqnarray}
H^{G}_{\Delta}&=& \sum_{ij}\Delta_{ij}\left(X^{\uparrow
0}_{i}X^{\downarrow 0}_{j} -X^{\downarrow 0}_{i}X^{\uparrow 0}_{j}
+H.c.\right)\nonumber\\&+&\mu\sum_iX_i^{00}. \label{1.5}
\end{eqnarray}
In contrast with the conventional real-space BCS Hamiltonian,
the strongly correlated BCS Hamiltonian given by Eq.(\ref{1.5})
is not an exactly solvable model. Since the Hubbard operators appear as the elements of the su$(2|1)$ superalgebra,
a natural framework to address this problem is provided by the
su$(2|1)$ coherent-state path-integral representation of the partition function.

\section{su$(2|1)$ coherent-state path integral representation of
the partition function}

In the su$(2|1)$ coherent-state basis the partition function
$$Z_{\Delta}=tr\ \exp (-\beta H^{G}_{\Delta})$$ takes the form of the
su$(2|1)$ coherent-space phase-space path integral (see Appendix
B):
\begin{equation}
Z_{\Delta}=\int D\mu (z,\xi )\ e^{S_{\Delta}}, \label{2.1}
\end{equation}%
where
$$D\mu (z,\xi )=\prod_{i,t}\frac{d\bar z_i(t)dz_i(t)}{2\pi
i(1+|z_i|^2)^2}\, d\bar\xi_i(t)d\xi_i(t).$$ Here $z_i$ is a
complex number that keeps track of the spin degrees of freedom,
while $\xi_i $ is a complex Grassmann parameter that describes the
charge degrees of freedom.

The effective action
\begin{eqnarray}
S_{\Delta}&=&i\sum_i\int_0^{\beta}a_i(t)dt-\sum_i\int_0^{\beta}\bar\xi_i
\left(\partial_t+ia_i\right)\xi_idt \nonumber\\&-&
\int_0^{\beta}H^{G,cl}_{\Delta}dt \label{2.2}\end{eqnarray} involves
the U(1)-valued connection one-form of the magnetic monopole
bundle (see Appendix A) that can formally be interpreted as a spin
"kinetic" term,
$$ia=-\langle z|\partial_t|z\rangle=\frac{1}{2}\frac{\dot{\bar z}z-\bar z\dot
z}{1+|z|^2},$$ with $|z\rangle$ being the su(2) coherent state.
This term is also frequently referred to as the Berry connection.
The dynamical part of the action takes the form
\begin{eqnarray}
H^{G,cl}_{\Delta}&=&\sum_{ij}\left(\Delta_{ij}\xi_i\xi_j \frac{\bar
z_j-\bar z_i}{\sqrt{(1+|z_i|^2)(1+|z_j|^2)}}
+H.c\right)\nonumber\\
&+&\mu \sum_i\bar\xi_i\xi_i. \label{2.3}
\end{eqnarray}
Here $z_i(t)$ and $\xi_i(t)$ are the dynamical fields. This
representation rigorously incorporates the constraint of no double
occupancy. Since the NDO constraint is explicitly resolved in representation (\ref{2.1}),
the dynamical variables $z_i$ and $\xi_i$ bear no local gauge redundancy
associated with the constraint-generated local gauge transformations, and
in contrast with the slave-particle fields,
are gauge independent.

Under the global SU(2) rotation,
\begin{eqnarray} z_i\rightarrow
\frac{uz_i+v}{-\overline{v}z_i+\overline{u}}, \quad \xi_i(t)\rightarrow
e^{i\psi_i }\xi_i,\quad a_i\to a_i+d\psi_i,
 \label{2.4}
\end{eqnarray}
where
\begin{eqnarray}
\psi_i=-i\log \sqrt{\frac{-v\overline{z}_i+u}{-\overline{v}z_i+\overline{%
u}}}, \quad \left(\begin{array}{ll}
u & v \\
-\overline{v} & \overline{u}%
\end{array}
\right) \in \mathrm{SU(2)}.
\end{eqnarray}
The effective action (\ref{2.2}) is invariant under the global spin rotations
given by Eqs. (\ref{2.4}).

Let us now make the following change of variables on the sublattice $B$,
\begin{equation}
z_i\to-\frac{1}{\bar z_i}, \quad \xi_i\to
\bar\xi_i\sqrt{\frac{z_i}{\bar z_i}}, \quad i\in B. \label{2.5}
\end{equation}
This transformation is equivalent to a SU(2) rotation (\ref{2.4})
with $u=0, v=1$  followed by a complex conjugation. Under this
transformation $\vec S_i\to -\vec S_i$ and the gauge potential
$a_i$ changes its sign, $a_i\to -a_i$. The effective action then
becomes
\begin{eqnarray}
S_{\Delta}&\to&S_{\Delta}= i\sum_{i\in
A}\int_0^{\beta}a_i(t)dt\nonumber-\mu N\beta/2\\&+&\sum_{i\in
A}\int_0^{\beta}\bar\xi_i \left(-\partial_t-ia_i-\mu\right)\xi_idt
\nonumber\\ &+&\sum_{i\in B}\int_0^{\beta}\bar\xi_i
\left(-\partial_t-ia_i+\mu\right)\xi_idt -
\int_0^{\beta}H^{G,cl}_{\Delta}dt \nonumber,\end{eqnarray} where
\begin{eqnarray}
H^{G,cl}_{\Delta}&=&\sum_{ij}\left(\Delta_{ij}\bar\xi_i\xi_j \langle
z_j|z_i\rangle +H.c\right),\nonumber\end{eqnarray} and $\langle
z_{i}|z_{j}\rangle$ stands for an inner product of the su(2)
coherent states, $$\langle z_{i}|z_{j}\rangle =\frac{1+\overline{z}%
_{i}z_{j}}{\sqrt{(1+|z_{j}|^{2})(1+|z_{i}|^{2})}}.$$ This can be
written in the form
\begin{eqnarray}
S_{\Delta}&=& i\sum_{i\in A}\int_0^{\beta}a_i(t)dt-\mu
N\beta/2\nonumber\\&+&\sum_{ij}\int_0^{\beta}\bar\xi_i(t)
G^{-1}_{ij}(t,s)\xi_j(s)dtds,\label{2.6}\end{eqnarray} where
$$G^{-1}_{ij}(t,s)=G^{-1}_{(0)ij}(t,s)-ia_i(t)\delta_{ij}\delta(t-s)+\Sigma_{ij}(t)\delta(t-s),$$
with $ \Sigma_{ij}= \Delta_{ij}\langle z_j|z_i\rangle$ and
$$G^{-1}_{(0)ij}(t,s)=\delta_{ij}(-\partial_{t}-\mu)\delta(t-s),\,
i\in A,$$
$$G^{-1}_{(0)ij}(t,s)=\delta_{ij}(-\partial_{t}+\mu)\delta(t-s),\,
i\in B.$$

The fermionic degrees of freedom in Eq.(\ref{2.1}) can formally be
integrated out to yield
$$ \int D\bar\xi
D\xi\exp{\left(\sum_{ij}\int_0^{\beta}\bar\xi_i(t)
G^{-1}_{ij}(t,s)\xi_j(s)dtds\right)} $$ $$=\exp{Tr\log G^{-1}}$$
\begin{equation}
=\exp{\left(Tr\log
G^{-1}_{(0)}+Tr\log(1-G_{(0)}ia+G_{(0)}\Sigma)\right)}.
\label{2.7}\end{equation} Here the trace has to be carried out
over both space and time indices. Calculating explicitly a
factor that comes from the zero order Green's function, we get
$$Z_0:=Z_{\Delta=a=0}=\exp{(Tr\log G^{-1}_{(0)}-\mu N\beta/2)}$$
$$=\exp{\left(\beta\sum_{i \in A}\log G^{-1}_{(0)}
+\beta\sum_{i \in B}\log G^{-1}_{(0)}-\mu N\beta/2\right)}$$
$$=\left(2\cosh\frac{\mu\beta}{2}\,e^{-\frac{\mu\beta}{2}}\right)^N,$$
which is a correct result for the partition function
of N noninteracting spinless fermions, $$Z_0=tr\,
e^{-\mu\int^{\beta}_0 \sum_if^{\dagger}_if_i}.$$

Up to this point no approximation has been made in the derivation
of the effective action. In fact, we are interested in a
derivation of an effective action  to describe a low-energy
dynamics of the spin degrees of freedom of the projected
strong-pairing Hamiltonian close to half-filling. For that
purpose we deduce an effective action in the spin degrees of
freedom by performing a perturbative expansion of the expression
$Tr\log(1-G_{(0)}ia+G_{(0)}\Sigma)$ in powers of $|\Delta|/\mu <<
1$. Physically, this corresponds to the lightly underdoped region of
the phase diagram. The second step consists in expanding the
obtained representation up to first order in $\partial_t$ and
second order in $\Delta_{ij}$ implying that eventually we will set
$i\to j$. This amounts to the so-called gradient expansion that
corresponds to the low-energy and long-wavelength limit of the
action. In this way we obtain
\begin{eqnarray}
&&Tr\log(1-G_{(0)}ia+G_{(0)}\Sigma)=\nonumber\\
&&-Tr(G_{(0)}ia)-\frac{1}{2}Tr(G_{(0)}\Sigma
G_{(0)}\Sigma).\label{2.8}\end{eqnarray} Note that
$Tr(G_{(0)}iaG_{(0)}\Sigma)=0$ since $\Sigma_{ii}=0$. This
expansion is justified in the limit $|\Delta|/\mu << 1, \,
\mu\beta>>1.$

The $a$-dependent term in Eq.(\ref{2.8}) contributes to the action
in the following way
$$-Tr(G_{(0)}ia)=-i\sum_{i\in
A}G_{(0)i}(0^{-})\int_0^{\beta}a_i(t)dt$$ $$-i\sum_{i\in
B}G_{(0)i}(0^{-})\int_0^{\beta}a_i(t)dt,$$ where
$G_{(0)i}(0^{-}):=\lim_{\epsilon \to 0}G_{(0)i}(-\epsilon), \,\,
\epsilon > 0$ and
\begin{equation}
G_{(0)i}^{A/B}(\tau)=\frac{e^{\mp\mu\tau}}{1+e^{\pm\mu\beta}}-\theta(\tau)e^{\mp\mu\tau}.
\label{2.9}
\end{equation}
Here the upper sign corresponds to the case $i\in A$, whereas the
lower one, to the case  $i\in B$. The explicit representation
(\ref{2.9})tells us that
\begin{equation}
\Delta S_1:=-Tr(G_{(0)}ia)=-i\sum_{i\in B}\int_0^{\beta}a_i(t)dt
+{\cal O}(e^{-\mu\beta}), \label{2.10}\end{equation} where it is
implied that $\mu\beta >>1$.

Let us now turn to the second term in Eq.(\ref{2.8}). We get
$$\Delta S_2:=-\frac{1}{2}Tr(G_{(0)}\Sigma G_{(0)}\Sigma)$$
$$=-\frac{1}{2}\sum_{ij}\!\!\int G_{(0)i}(t_1-t_2)\Sigma_{ij}(t_2)
G_{(0)j}(t_2-t_1)\Sigma_{ji}(t_1)dt_1dt_2.$$ Introducing new
variables , $\tau=\frac{t_1-t_2}{2}, \, \eta=\frac{t_1+t_2}{2}$,
and expanding the product $\Sigma_{ij}(\eta +\tau)\Sigma_{ji}(\eta
-\tau)=\Sigma_{ij}(\eta)\Sigma_{ji}(\eta)+{\cal O}(\tau)$ (this
corresponds to the gradient expansion in imaginary time \cite{ranninger}), gives to
the lowest order
\begin{equation}
\Delta
S_2=-\frac{1}{2}\sum_{ij}\int^{\beta}_{-\beta}G^A_{(0)}(\tau)G^B_{(0)}(-\tau)d\tau
\!\!\int^{\beta}_0\Sigma_{ij}(\eta)\Sigma_{ji}(\eta)d\eta
\label{2.11}\end{equation} With the help of Eqs.(\ref{2.9}) we get
\begin{equation}
-\frac{1}{2}\int^{\beta}_{-\beta}G^A_{(0)}(\tau)G^B_{(0)}(-\tau)d\tau
=\frac{1}{2\mu}(1+{\cal
O}(e^{-\mu\beta})).\label{2.12}\end{equation}

The effective spin action is given by the sum of all the term
evaluated above:
\begin{equation}
Z^{eff}_{\Delta}/Z_0=\int D\mu (z,\bar z)\ e^{S^{eff}_{\Delta}},
\label{2.13}
\end{equation}%
where the SU(2) invariant measure factor
$$D\mu (z,\bar z)=\prod_{i,t}\frac{d\bar z_i(t)dz_i(t)}{2\pi
i(1+|z_i|^2)^2}$$ and
\begin{eqnarray}
S_{\Delta}^{eff}&=&i\sum_{i\in A}\int_0^{\beta}a_i(t)dt-i\sum_{i\in
B}\int_0^{\beta}a_i(t)dt\nonumber\\
&+&\sum_{ij}\int_0^{\beta}\frac{|\Delta_{ij}|^2}{2\mu}|\langle
z_i|z_j\rangle|^2dt.\label{2.14}\end{eqnarray}

Let us now rotate the spin on the sublattice $B$ back to their
initial position, $z_i\to-1/\bar z_i,$. Under this transformation
$$|\langle z_i|z_j\rangle|^2 \to 1-|\langle z_i|z_j\rangle|^2,
\quad a_{i\in B}\to -a_{i\in B}.$$
In this way we finally get,
\begin{eqnarray}
S_{\Delta}^{eff}&=&i\sum_{i}\int_0^{\beta}a_i(t)dt\nonumber\\
&-&\sum_{ij}J^{(U=\infty,\mu)}_{ij}\int_0^{\beta}\left(|\langle
z_i|z_j\rangle|^2-1\right)dt,\label{2.15}\end{eqnarray} where the
long-wavelength limit $(j\to i)$ is implied. This action describes the
antiferromagnetic Heisenberg model with the effective coupling (see Appendix A)
\begin{equation}
J^{(U=\infty,\mu)}_{ij}=|\Delta_{ij}|^2/2\mu>0.
\label{j}\end{equation}
The resulting model is the low-energy action for the fully projected $(U=\infty)$ BCS Hamiltonian in the vicinity of half filling ($\mu $ is large but finite).
In the explicit low-energy and
long-wavelength limit the $2D$ quantum action (\ref{2.15}) reduces to that of
the $3D$ classical nonlinear sigma-model (see Appendix C). Taking into consideration the RG analysis of that sigma-model action \cite{auerbach95} it then follows that 
the ground state of the $2D$ Gutzwiller-projected BCS Hamiltonian is AF magnetically ordered at sufficiently
low doping.

Right at half filling $\mu\to\infty$, producing in this case $J^{(U=\infty,\mu)}_{ij}\to 0$. 
However, this does not contradict
Park's observation that the projected BCS Hamiltonian (for large but finite $U$) possesses a long-range AF ordered ground state right at half filling \cite{park}.
This can be seen as follows.
Essentially, we are interested in the Hamiltonian
\begin{equation}
H_{\Delta+U}=\sum_{ij}\Delta_{ij}\left(c_{i\uparrow}^{\dagger}c^{\dagger}_{j\downarrow}
-c_{i\downarrow}^{\dagger}c^{\dagger}_{j\uparrow} +H.c.\right)+U\sum_in_{i\uparrow}n_{i\downarrow},
\label{p1}
\end{equation}
which, in the $U\to\infty$ limit, reduces to $H^{G}_{\Delta}$ as in Eq. (\ref{1.5}).
We now back off from the infinite $U$ limit, the effects of double occupancy need to be build in perturbatively in powers of $\Delta/U$.

Let us make the following unitary transformation of the electron operators for all sites $j\in B$:
$$c_{j\uparrow}\to c^{\dagger}_{j\downarrow}, \quad c_{j\downarrow}\to -c^{\dagger}_{j\uparrow}.$$
In this way we get
\begin{eqnarray}
H_{\Delta+U}&\to&
-\sum_{ij}\Delta_{ij}\left(c_{i\sigma}^{\dagger}c_{j\sigma}
 +H.c.\right)\nonumber\\&+&U\sum_in_{i\uparrow}n_{i\downarrow}-U\sum_{i\in B}n_i.
\label{p2}
\end{eqnarray}
Using the representation
$$n_{i\uparrow}n_{i\downarrow}=-\frac{2}{3}\vec Q_i^2+\frac{n_i}{2},$$
where $\vec Q_i$ is the electron spin operator, we can write down the partition function as
\begin{equation}
Z_{\Delta+U}=\int D\vec\phi D\bar\Psi D\Psi \exp{\int_0^{\beta}{\cal L}_{\Delta+U}dt},
\end{equation}
with
$${\cal L}_{\Delta+U}=\frac{-3U}{8}\sum_i\vec\phi_i^2-U\sum_i\bar\Psi_i\vec\tau\vec\phi_i\Psi_i$$
$$+\sum_{ij}\Delta_{ij}\left(\bar\Psi_i\Psi_j+H.c\right)+U\sum_{i\in B}\bar\Psi_i\Psi_i ,$$
$$\Psi_i=(c_{i\uparrow}, c_{i\downarrow})^t.$$

For $U>>\Delta$ one gets $\vec\phi_i^2\approx 1$. As a result, in this limit, one can make the identification
$\vec\phi_i=2\vec S_i^{cl}(\bar z,z)$ \cite{nagaosa}. Using the identity 
$$2\vec S^{cl}\vec \tau = V\tau_zV^{\dagger},$$ where
\begin{eqnarray}
V=\frac{1}{\sqrt{1+|z|^2}}\left(\begin{array}{ll}
1 &-\bar z \\
z & \,\,\,1%
\end{array}\right).
\end{eqnarray}
and rotating now the spinors to the $z$ axis, $\Psi\to V\Psi,$ we get
$${\cal L}_{\Delta+U}\to \sum_i\bar \Psi_i\left(-\partial_t -U\tau_z\right)\Psi_i +U\sum_{i\in B}\bar\Psi_i\Psi_i$$
$$+\sum_i\bar\Psi_iV^{\dagger}_i(-\partial_tV_j)\Psi_j+\sum_{ij}\Delta_{ij}\bar\Psi_iV^{\dagger}_iV_j\Psi_j .$$
The fermionic degrees of freedom can now be integrated out in the low-energy limit,  yielding
\begin{equation}
Z_{\Delta+U}/Z_0=\int D\mu (z,\bar z)\ e^{S^{eff}_{\Delta+U}},
\label{p3}
\end{equation}%
where
$$D\mu (z,\bar z)=\prod_{i,t}\frac{d\bar z_i(t)dz_i(t)}{2\pi
i(1+|z_i|^2)^2}$$ and the effective low-energy action is again given by Eq. (\ref{2.15}) but now with $J^{(U,\mu=\infty)}_{ij}=|\Delta_{ij}|^2/2U>0.$
This consideration provides an independent proof of the equivalence of the low-energy physics of the $2D$ Heisenberg AF model and  the Gutzwiller-projected strong-pairing Hamiltonian at half filling first established in \cite{park}, though within quite a different approach.

To consider both cases  simultaneously one should start directly with the Hamiltonian
\begin{eqnarray}
H_{\Delta+U+\mu}&=&\sum_{ij}\Delta_{ij}\left(c_{i\uparrow}^{\dagger}c^{\dagger}_{j\downarrow}
-c_{i\downarrow}^{\dagger}c^{\dagger}_{j\uparrow} +H.c.\right)\nonumber \\&+&U\sum_in_{i\uparrow}n_{i\downarrow}
+\mu\sum_i(1-n_i)^2.
\label{full}
\end{eqnarray}
The low-energy AF action in this case is supposed to be specified by the exchange coupling constant $J_{ij}^{(U,\mu)}$. Since a general case of large but finite values of the parameters $U$ and $\mu$ is technically involved, we consider two limiting cases, namely, that of $U=\infty$, and  $\mu $ is large but finite and that of $\mu=\infty$, and $U$ is large but finite.

Physically, for the fully Gutzwiller-projected ($U=\infty)$ BCS Hamiltonian the AF order sets in at low doping due to the superexchange that involves empty states. In the second case when $\mu$ tends to $\infty$ and
the repulsion parameter $U$ is large but finite, there is a small probability for the doubly occupied electron states to exist, which results in the emergence of the long-range magnetic order through the virtual superexchange process that involves the doubly occupied states.  In a real physical system both mechanisms are evidently at work which accounts for the emergence of the AF phase of the cuprate superconductors close to half filling.

\section{Second-order operator perturbation theory}

In this section we briefly comment on another derivation of the low-energy representation of the projected BCS Hamiltonian now following a more conventional operator approach.

Let us rewrite the Hamiltonian given in Eq. (\ref{1.5}) in the following way:
\begin{equation}
H^G_{\Delta}=H_0+V
\end{equation}
with
\begin{equation}
H_0=\mu\sum_iX_i^{00}
\end{equation}
\begin{equation}
V=\sum_{ij}\Delta_{ij}\left(X^{\uparrow
0}_{i}X^{\downarrow 0}_{j} -X^{\downarrow 0}_{i}X^{\uparrow 0}_{j}
+H.c.\right)
\end{equation}
At half-filling, we take $\mu \rightarrow \infty$. This ensures that any state with a finite number of holes is projected out from the theory. Close to half-filling, $\mu$ is large and therefore we can treat $V$ as a perturbation to $H_0$. The ground state of $H_0$ contains no holes and is highly degenerate, corresponding to all possible spin orientations in the half-filled limit. We denote this manifold by $|0_g \rangle$.

Let us now define the operator $P_0$ that projects into the subspace with no holes, that is, it projects into the ground state of $H_0$. Up to second order in $V$, we can define the effective Hamiltonian \cite{messiah99}:
\begin{equation}
H^{gr}_{eff}= P_0 V P_0 +\sum_{\phi_n \neq O_g} \frac{P_0 V |\phi_n \rangle \langle \phi_n | V P_0}{\epsilon_0-\epsilon_n}
\end{equation}
where $\epsilon_0=0$ is the ground state energy and $|\phi_n \rangle$ is an eigenstate of $H_0$ with eigenvalue $\epsilon_n$.

Since $V$ does not conserve the number of holes, the first order contribution is zero, that is, $P_0 V P_0 = 0$. In the next order, we have to calculate matrix elements like $\langle \phi_n |V|0_g \rangle$. They are non-zero only if $|\phi_n \rangle$ is a state containing two holes. In other words, the second order term is related to virtual transitions where two neighboring holes are first created and then destroyed, that is, Cooper pairing fluctuations in the system. It is therefore clear that $\epsilon_0-\epsilon_n=-2\mu$ and that our effective Hamiltonian now bwcomes 
\begin{eqnarray}
H^{gr}_{eff} &=& - \sum_{ij} \frac{|\Delta_{ij}|^2}{2\mu} (X^{\uparrow 0}_{i}X^{\downarrow 0}_{j}-X^{\downarrow 0}_{i}X^{\uparrow 0}_{j})\nonumber\\
&\times&(X^{0 \downarrow}_{j}X^{0 \uparrow}_{i}-X^{0 \uparrow}_{j}X^{0 \downarrow}_{i})
\end{eqnarray}

At half filling,
\begin{equation}
X^{\uparrow \downarrow}_{i}=S_i^+,\quad
X^{\downarrow \uparrow}_{i}=S_i^-,
\end{equation}
\begin{equation}
X^{\uparrow \uparrow}_{i}-X^{\downarrow \downarrow}_{i}=2S_i^z,\quad
X^{\uparrow \uparrow}_{i}+X^{\downarrow \downarrow}_{i}=\tilde{n}_i=1,
\end{equation}
and after a straightforward algebra we get
\begin{equation}
H^{gr}_{eff} =  \sum_{ij} \frac{|\Delta_{ij}|^2}{2\mu} \left( \vec S_i \vec S_j - \frac{1}{4}
\right).
\end{equation}

As a result, using a simple perturbative scheme, we find that the ground-state of the strong-pairing Gutzwiller-projected BCS Hamiltonian is indeed identical to that of the antiferromagnetic Heisenberg model with coupling given by $J_{ij}=\frac{|\Delta_{ij}|^2}{2\mu}$, close to half-filling.

\section{Conclusion}

We conclude by discussing the physical implications of the close connection between the Gutzwiller-projected BCS Hamiltonian and the t-J model of the high-T$_c$ superconductors. Our result shows that the ground state of the Gutzwiller-projected BCS Hamiltonian can in principle be considered a reference state of a lightly doped Mott insulator. Since the Gutzwiller projection does not commute with the BCS Hamiltonian \cite{park}, this state does not coincide with the Gutzwiller-projected BCS ground state which is just the short-range RVB state proposed by Anderson \cite{anders}.

The RVB state is known to show no long-range order even at half filling. In contrast,
right at half filling as well as in the immediate vicinity of half filling the ground state of the strongly correlated BCS Hamiltonian  exhibits long-ranged AF order as is observed in the cuprate superconductors.
Note also that the low-energy action that corresponds to the strong-pairing projected BCS Hamiltonian cannot in itself account for the weakening as well as the eventual disappearance of the magnetic ordering as the hole concentration increases. This effect is produced by the growing influence of the kinetic t-term that gradually destroys the long-range ordered state. Therefore, one needs to include the kinetic t-term into consideration to regain the full Gutzwiller-projected BCS Hamiltonian, $H^G_{BCS}$ given in Eq. (\ref{1.1}),
in order to be able to describe the actual behavior of the high-T$_c$ phase diagram away from half filling.

At a moderate, non-zero doping the RVB wavefunction and its improvements \cite{DB} yield good agreement with experiments \cite{18} as well as with numerical studies \cite{sorella02} and are conjectured to be
a good ansatz wavefunctions for the t-J model in this region \cite{gros}. In
doped regimes sufficiently away from half filling, the RVB state turns out to be indeed qualitatively similar to the ground state of $H^G_{BCS}$ \cite{park}. One can therefore
conclude that the ground-state wavefunction of the $H^G_{BCS}$ Hamiltonian appears as a natural generalization of Anderson's RVB state for low doping.

\appendix

\section{su(2) algebra and coherent states}

Consider the su(2) algebra in the lowest $s=1/2$ representation:
\begin{equation}
[S_z,S_{\pm}]=\pm S_{\pm},\quad [S_{+},S_{-}]=2S_z, \quad \vec
S^2=3/4.\label{1a.1}\end{equation} Acting with the ``lowering``
spin operator $S^{-}$ on the ``highest weight`` state
$|\uparrow\rangle$ we get the normalized su(2) CS parametrized by
a complex number $z$
\begin{equation}
|z\rangle=\frac{1}{\sqrt{1+|z|^2}}\exp( zS^{-})|\uparrow\rangle=
\frac{1}{\sqrt{1+|z|^2}}(|\uparrow\rangle +z|\downarrow\rangle).
\label{1a.2}\end{equation} In the basis spanned by the vectors
$|\uparrow\rangle, \,|\downarrow\rangle$ we have
$S_{+}=|\uparrow\rangle|\langle\downarrow|,
\,S_{-}=|\downarrow\rangle|\langle\uparrow|, \,
S_z=\frac{1}{2}(|\uparrow\rangle|\langle\uparrow|-|\downarrow\rangle|\langle\downarrow|).$
The CS symbols of the su(2) generators are then easily evaluated
to be ($S^{cl}:=\langle z|S|z\rangle$):
\begin{eqnarray}
S_{+}^{cl}:&=&\frac{z}{1+|z|^2},\quad
S_{-}^{cl}=\frac{\bar z}{1+|z|^2},\nonumber\\
S_z^{cl}&=&\frac{1}{2}\frac{1-|z|^2}{1+|z|^2},\quad \vec
S_{cl}^2=1/4, \,(\vec S^2)_{cl}=3/4.\label{1a.3}\end{eqnarray}

There is a one-to-one correspondence between the su(2) generators
~(\ref{1a.1}) and their CS (classical) symbols given by Eqs.
~(\ref{1a.3}). Given a quantum Hamiltonian $H=H(\vec S)$, the corresponding
imaginary time phase-space action takes on the form,
\begin{equation}
{\cal A}_{su(2)}(\bar z,z)=-\,\int^{\beta}_0\langle z|
\frac{d}{dt}+H|z\rangle dt, \label{1a.4}\end{equation} with the
kinetic term being given by
$$ia=-\langle z|\frac{d}{dt}|z\rangle= \frac{1}{2}\frac{\dot{\bar z}z-\bar z\dot
z}{1+|z|^2}.$$
In particular, for the quantum $s=1/2$  Heisenberg model,
$$H=J\sum_{ij}(\vec S_i\vec S_j-1/4),$$ one gets
$$H^{cl}=\frac{J}{2}\sum_{ij}(|\langle z_i|z_j\rangle|^2-1).$$

From the geometrical viewpoint, the su(2) coherent states
$|z\rangle$ can be thought of as sections of the magnetic monopole
bundle $P(S^2, U(1))$, with the U(1) connection one-form, $ia$,
frequently refereed to as the Berry connection. Base space of that
bundle, two-sphere $S^2$, appears as a classical phase-space of
spin, whereas its covariantly constant sections, $|z\rangle:\,
\,(\partial_t+ia)|z\rangle=0$, form a Hilbert space of a quantum spin.

\section{su$(2|1)$ superalgebra and coherent states}

Acting with the ``lowering`` superspin operators
$X^{\downarrow\uparrow}$ and $X^{\downarrow 0}$ on the ``highest
weight`` state $|\uparrow\rangle$ we get the normalized  su$(2|1)$
coherent state in the 3D fundamental representation,
\begin{eqnarray}
|z,\xi\rangle&=&(1+\bar{z}z
+\bar{\xi}\xi)^{-1/2}\exp\left(zX^{\downarrow\uparrow}+\xi
X^{0\uparrow}\right)|\uparrow\rangle \nonumber\\
&=&(1+\bar{z}z +\bar{\xi}\xi)^{-1/2}(|\uparrow\rangle
+z|\downarrow\rangle+\xi |0\rangle), \label{1b.1}
\end{eqnarray} where $z$ is a complex number, and $\xi $ is a complex
Grassmann parameter. The Grassmann parameter appears here due to
the fact that $X^{\downarrow 0}$ is a fermionic operator in
contrast with the operator $X^{\downarrow\uparrow}$. The product
$\xi X^{0\uparrow}$ represents therefore a bosonic quantity as
required.

At $\xi =0$, the su$(2|1)$ CS reduces to the ordinary su(2) CS,
$|z,\xi=0\rangle\equiv |z\rangle$ ~(\ref{1a.2}), parametrized by a
complex coordinate $z \in$ CP$^1\simeq S^2$. In contrast, at $z=0$, it
represents a pure fermionic CS.

The CS symbols of the $X$ operators, $X_{cl}:=\langle
z,\xi|X|z,\xi\rangle$, are
\begin{eqnarray}
X^{0\downarrow}_{cl}&=&-\frac{z\bar\xi}{1+|z|^2},\quad
X^{\downarrow0}_{cl}=-\frac{\bar
z\xi}{1+|z|^2},\nonumber\\
X^{0\uparrow}_{cl}&=&-\frac{\bar\xi}{1+|z|^2},\quad X^{\uparrow
0}_{cl}=-\frac{
\xi}{1+|z|^2},\nonumber\\
Q^{+}_{cl}=X^{\uparrow\downarrow}_{cl}&=&\frac{z}{1+|z|^2}\left(1-\frac{\bar\xi\xi}{1+|z|^2}\right),\nonumber\\
Q^{-}_{cl}=X^{\downarrow\uparrow}_{cl}&=&\frac{\bar
z}{1+|z|^2}\left(1-\frac{\bar\xi\xi}{1+|z|^2}\right),\nonumber\\
Q^z_{cl}=\frac{1}{2}(X^{\uparrow\uparrow}_{cl}-X^{\downarrow\downarrow}_{cl})&=&
\frac{1}{2}\frac{1-|z|^2}{1+|z|^2}\left(1-\frac{\bar\xi\xi}{1+|z|^2}\right).
\label{1b.2}\end{eqnarray} Given a Hamiltonian as a polinomial
function of the Hubbard operators,$H=H(X)$, the corresponding
imaginary time phase-space action takes on the form,
\begin{equation}
{\cal A}_{su(2|1)}=-\,\int^{\beta}_0\langle z,\xi|
\frac{d}{dt}+H(X)|z,\xi\rangle dt, \label{1b.3}\end{equation} with
the kinetic term given by
\begin{eqnarray}
\langle
z,\xi|(-\frac{d}{dt})|z,\xi\rangle=\frac{1}{2}\frac{\dot{\bar
z}z-\bar z\dot z
+\dot{\bar\xi}\xi-\bar\xi\dot\xi}{1+|z|^2+\bar\xi\xi}.
\label{1b.4}\end{eqnarray}  Substituting  $H(X)=H_{\Delta}$ into
Eq.(\ref{1b.3}) and making the change of variables $z_i\to z_i\,
\, \xi_i\to\xi_i\sqrt{1+|z_i|^2}$, we are led to the effective
action (\ref{2.2}).

\section{Nonlinear $\sigma$-model}

Consider the $1D$ $s$-spin quantum AF Heisenberg model on a
bipartite lattice, $L=A\oplus B$,
\begin{equation}
H=\sum_{<ij>}J_{ij}\left(\vec S_i\vec S_j-s^2\right),\,\,
J_{ij}>0, \label{c.1}\end{equation} where $J_{ij}=J$ for the nn
sites and $J_{ij}=0$ otherwise. Let us make the change $J\to J/2s$
and consider $H^{cl}=2s H^{cl}_{s=1/2}$. The coherent-state action
turns out to be $\propto 2s$, $S=(2s)\,S_{s=1/2},$ where
\begin{equation}
H^{cl}_{s=1/2} = \sum_{<ij>}\frac{J_{ij}}{2} \left(|\langle
z_i|z_j\rangle|^2-1\right), \label{c.2}\end{equation} so that
$$S_{s=1/2}=i\sum_{i\in}\int_0^{\beta}a_i(t)dt-\int^{\beta}_0H^{cl}_{s=1/2}$$
coincides with the action given by Eq.(\ref{2.15}), provided we
identity $J_{ij}$ in Eq.(\ref{c.2}) with $|\Delta_{ij}|^2/\mu.$

To proceed, notice the following identity
\begin{equation}
|\langle z_i|z_j\rangle|^2 = \exp\Phi_{ij},\quad
 \label{eq:c.3}\end{equation} where
\begin{equation}
\Phi_{ij}= F(\bar z_i,z_j)+ F(\bar z_j,z_i)- F(\bar
z_i,z_i)-F(\bar z_j,z_j)\le 0 \label{c.4}\end{equation} and
$F(\bar z_i,z_j)=\log(1+\bar z\_iz_j)$ is the so-called SU(2)
Kaehler potential, in terms of which the $\sigma$-model action can
be derived.

In order to obtain the $N\acute{e}el$ ground state, we should have
$\vec S^{cl}_A=- \vec S^{cl}_B$. To this end, let us make the
following change of variables in the path integral \cite{kochetov}:
$$z_i\to z_i+\xi_i,\quad i\in A;\quad z_i\to
-1/(\bar z_i-\bar\xi_i) \quad i\in B,$$ where $\xi_i,\,\bar\xi_i$
stand for a set of auxiliary fields $\sim a.$ In this way we get
\begin{equation}
H^{cl}_{s=1/2}=\frac{J}{2} \sum_{i}F_{\bar
z_iz_i}\frac{\partial\bar z}{\partial x_i} \frac{\partial
z}{\partial x_i}a^2 +2J\sum_{i}F_{\bar z_iz_i}\bar\xi_i\xi_i,
\label{c.4}\end{equation} where we have put $z_i=z(x_i),\quad
z_{i+1}=z(x_i+a)$, with $a$ being a lattice spacing. The total
action becomes,
\begin{eqnarray}
&&S_{s=1/2}= S_B\nonumber\\
&&+\int dt\sum_{i}F_{\bar z_iz_i}\left(\xi_i\dot{\bar
z}_i-\bar\xi_i\dot
z_i-2J\bar\xi_i\xi_i-\frac{J}{2}\frac{\partial\bar z} {\partial
x_i}\frac{\partial z}{\partial x_i}a^2 \right)\nonumber
\end{eqnarray}
where $S_B$ is the Berry phase term which will be considered
shortly. The auxiliary fields $\bar\xi_i$ and $\xi_i$ can be
eliminated to yield
$$ S_{s=1/2}=  \int dt\sum_{i}F_{\bar z_iz_i}\left\{-\frac{J}{2}
\frac{\partial\bar z}{\partial x_i}\frac{\partial z}{\partial
x_i}a^2 -\frac{1}{2J}\dot{\bar z}_i\dot z_i\right\}+S_B.$$
Restoring an explicit $s$-dependence and going over to the
continuum ($a\to 0$) limit, finally yields
\begin{equation}
S_{AF}=-\frac{1}{g^2}\int dxdt\,F_{\bar zz}\left(c\partial_x\bar z
\partial_xz +c^{-1}\dot{\bar z}\dot z\right) +S_B,
\label{c.5}\end{equation} where $c=2Jsa$ is the spin wave
velocity, $g^2=1/s$ is the coupling of the $\sigma$-model. We are
free to choose units so that $c=1$ and the action becomes Lorentz
invariant:
\begin{eqnarray}
S_{AF}&=&-\frac{1}{g^2}\int dxdt\,(g_{\bar zz}\partial_{\mu}\bar z
\partial_{\mu}z)\nonumber\\
&=&-\frac{1}{g^2}\int dxdt\,\frac{\partial_{\mu}\bar z
\partial_{\mu}z} {(1+|z|^2)^2}, \qquad \mu=0,1.
\label{c.6}\end{eqnarray} The generalization of this result to the
case of the $D$-dimensional quantum antiferromagnet is trivial: in
the low-energy quasiclassical (large spin $s$) limit it is
described by the classical $D+1$ dimensional $\sigma$-model (\ref{c.6})
where $\mu=0,1,2..,D.$

The Berry phase term in $1D$ becomes
$$S_B=\frac{i}{2}\int_{S^2} \,da=s\int_{S^2}\frac{dz\wedge d\bar z}{(1+|z|^2)^2}=2\pi
isN,$$ Where $N$ is an integer, the Brouwer degree of the map
$z(x,t):\, S^2\to S^2.$ Thus, in $1D$ this phase term turns into a
topological (metric independent) invariant that gives rise to
dramatical consequences on ground state degeneracy and low-energy
spectrum. In higher dimensions the Berry phase term does not
contribute to the action.


\begin{thebibliography}{99}







\bibitem{dagotto94}
E. Dagotto, Rev. Mod. Phys. {\bf 66}, 763 (1994).



\bibitem{lee06}
P.~A. Lee, N. Nagaosa, and X-G. Wen, Rev. Mod. Phys. {\bf 78}, 17 (2006).



\bibitem{sorella02}
S. Sorella \emph{et al.}, Phys. Rev. Lett. {\bf 88}, 17002 (2002).



\bibitem{park}
K. Park, Phys. Rev. Lett. {\bf 95}, 027001 (2005);
Phys. Rev. B {\bf 72}, 245116 (2005).

\bibitem{auerbach95}
A. Auerbach, "Interacting Electrons and Quantum Magnetism" (Springer-Verlag, NY: 1995).

\bibitem{ranninger} M. Cuoco and J. Ranninger, Phys. Rev. B {\bf 70}, 104509 (2004).


\bibitem{nagaosa} N. Nagaosa " Quantum Field Theory in Strongly Correlated Electronic Systems"
(Springer: 1998).

\bibitem{messiah99}
A. Messiah, "Quantum Mechanics" (Dover Publications, Mineola, NY: 1999).

\bibitem{18}
A. Paramekanti, M. Randeiria, and N. Trivedi, Phys. Rev. Lett. {\bf 87}, 217002 (2001).

\bibitem{gros} F.C. Zhang, C. Gros, T.M. Rice, and H.Shiba, Supercond. Sci. and Tech., {\bf 1}, 36 (1988);

\bibitem{kochetov} E.A. Kochetov, unpublished.

\bibitem{anders} P.W. Anderson, Science {\bf 235}, 1196 (1987).

\bibitem{DB} D. Eichenberger and D. Baeriswyl, cond-mat 0708.2795, accepted in Phys. Rev. B.
\end{thebibliography}
\end{document}